\documentclass[12pt]{article}
\usepackage{amsmath,amsfonts,amssymb}

\begin{document}
\thispagestyle{empty}

\def\theequation{\arabic{section}.\arabic{equation}}
\def\a{\alpha}
\def\b{\beta}
\def\g{\gamma}
\def\d{\delta}
\def\dd{\rm d}
\def\e{\epsilon}
\def\ve{\varepsilon}
\def\z{\zeta}
\def\B{\mbox{\bf B}}\def\cp{\mathbb {CP}^3}

\newcommand{\h}{\hspace{0.5cm}}

\begin{titlepage}
\vspace*{1.cm}
\renewcommand{\thefootnote}{\fnsymbol{footnote}}
\begin{center}
{\Large \bf Three-point Correlation Function of Giant Magnons} \vskip 0cm
{\Large\bf in the Lunin-Maldacena background}
\end{center}
\vskip 1.2cm \centerline{\bf Changrim  Ahn and Plamen Bozhilov
\footnote{On leave from Institute for Nuclear Research and Nuclear
Energy, Bulgarian Academy of Sciences, Bulgaria.}}

\vskip 10mm

\centerline{\sl Department of Physics} \centerline{\sl Ewha Womans
University} \centerline{\sl DaeHyun 11-1, Seoul 120-750, S. Korea}
\vspace*{0.6cm} \centerline{\tt ahn@ewha.ac.kr,
bozhilov@inrne.bas.bg}

\vskip 20mm

\baselineskip 18pt

\begin{center}
{\bf Abstract}
\end{center}

We compute semiclassical three-point correlation function, or structure
constant, of two finite-size (dyonic) giant magnon string states and a
light dilaton mode in the Lunin-Maldacena background, which 
is the $\gamma$-deformed, or $TsT$-transformed $AdS_5\times
S_{\gamma}^5$, dual to $\mathcal{N} = 1$ super Yang-Mills theory. We also
prove that an important relation between the structure constant and
the conformal dimension, checked for the $\mathcal{N} = 4$ super
Yang-Mills case, still holds for the $\gamma$-deformed string
background.

\end{titlepage}
\newpage
\baselineskip 18pt

\def\nn{\nonumber}
\def\tr{{\rm tr}\,}
\def\p{\partial}
\newcommand{\non}{\nonumber}
\newcommand{\bea}{\begin{eqnarray}}
\newcommand{\eea}{\end{eqnarray}}
\newcommand{\bde}{{\bf e}}
\renewcommand{\thefootnote}{\fnsymbol{footnote}}
\newcommand{\be}{\begin{eqnarray}}
\newcommand{\ee}{\end{eqnarray}}

\vskip 0cm

\renewcommand{\thefootnote}{\arabic{footnote}}
\setcounter{footnote}{0}

\setcounter{equation}{0}
\section{Introduction}

As is well known, the correlation functions of any conformal field
theory can be determined  in principle in terms of the basic
conformal data $\{\Delta_i,C_{ijk}\}$, where $\Delta_i$ are the
conformal dimensions defined by the two-point correlation functions
\begin{equation}
\left\langle{\cal O}^{\dagger}_i(x_1){\cal O}_j(x_2)\right\rangle=
\frac{C_{12}\delta_{ij}}{|x_1-x_2|^{2\Delta_i}}
\end{equation}
and $C_{ijk}$ are the structure constants in the OPE
\begin{equation}
\left\langle{\cal O}_i(x_1){\cal O}_j(x_2){\cal
O}_k(x_3)\right\rangle=
\frac{C_{ijk}}{|x_1-x_2|^{\Delta_1+\Delta_2-\Delta_3}
|x_1-x_3|^{\Delta_1+\Delta_3-\Delta_2}|x_2-x_3|^{\Delta_2+\Delta_3-\Delta_1}}.
\end{equation}
Thus, the determination of the initial conformal data for a given
CFT is the most important step in the conformal bootstrap approach.

In view of the AdS/CFT duality \cite{AdS/CFT}, between strings on
$AdS_5\times S^5$ and $\mathcal{N}=4$ super Yang-Mills (SYM) theory, the
correlators of single-trace conformal primary operators on the gauge
theory side, in the planar limit, should be related to the
correlation functions of the corresponding closed-string vertex
operators. The conformal dimension $\Delta$ can be expressed in
terms of the conserved charges and the string tension by the
marginality condition on the vertex operator.

In particular, the three-point functions of two heavy operators and
a light dilaton operator can be approximated by a supergravity
vertex operator evaluated at the heavy classical string
configuration: \bea \nn \langle
V_{H}(x_1)V_{H}(x_2)V_{L}(x_3)\rangle=V_L(x_3)_{\rm classical}.
\eea For $\vert x_1\vert=\vert x_2\vert=1$, $x_3=0$, the correlation
function reduces to \bea \nn \langle
V_{H}(x_1)V_{H}(x_2)V_{L}(0)\rangle=\frac{C_{123}}{\vert
x_1-x_2\vert^{2\Delta_{H}}}. \eea Then, the normalized structure
constants \bea \nn \mathcal{C}_3=\frac{C_{123}}{C_{12}} \eea can be
found from \bea \label{nsc} \mathcal{C}_3=c_{\Delta}V_L(0)_{\rm
classical}, \eea where $c_{\Delta}$ is the normalized constant of
the corresponding light vertex operator.

Recently, there has been an impressive progress in the semiclassical
calculations of two, three, and four-point functions with two heavy
operators \cite{Janik:2010gc}-\cite{AR1106}. Almost all of these
achievements are in the framework of the duality between string
theory in $AdS_5\times S^5$ and $\mathcal{N}=4$ SYM. An exception is
the paper \cite{AR1106}, considering the case of strings on
Lunin-Maldacena background \cite{LM05}, dual to $\mathcal{N}=1$ SYM
in four dimensions. In particular, the three-point correlation
function of two infinite-size giant magnons and the dilaton has been
obtained there. Our aim here is to generalize this result to the
case of {\it finite-size} dyonic giant magnons.

\setcounter{equation}{0}
\section{Three-point correlation function}

The bosonic part of the Green-Schwarz action for strings on the
$\gamma_i$-deformed $AdS_5\times S_{\gamma_i}^5$ \cite{AAF05}
reduced to $R_t\times S_{\gamma_i}^5$ can be written as (the common
radius $R$ of $AdS_5$ and $S_{\gamma_i}^5$ is set to 1)
\bea\label{BGS} S&=&-\frac{T}{2}\int d\tau
d\sigma\left\{\sqrt{-\gamma}\gamma^{ab}\left[-\p_a t\p_b t+\p_a
r_i\p_br_i+Gr_i^2\p_a\phi_i\p_b\phi_i \right.
\right.\\ \nn && +\left.\left.Gr_1^2r_2^2r_3^2
\left(\tilde{\gamma}_i\p_a\phi_i\right)
\left(\tilde{\gamma}_j\p_b\phi_j\right) \right]\right.\\ \nn
&&-2G\left.\epsilon^{ab}\left(\tilde{\gamma}_3r_1^2r_2^2\p_a\phi_1\p_b\phi_2
+\tilde{\gamma}_1r_2^2r_3^2\p_a\phi_2\p_b\phi_3
+\tilde{\gamma}_2r_3^2r_1^2\p_a\phi_3\p_b\phi_1\right)\right\} ,\eea
where $T$ is the string tension, $\gamma^{ab}$ is the worldsheet
metric, $\phi_i$  are the three isometry angles of the deformed
$S_{\gamma_i}^5$, and \bea\label{roG} \sum_{i=1}^{3}r_i^2=1,\h
G^{-1}=1+\tilde{\gamma}_3^2r_1^2r_2^2+\tilde{\gamma}_1^2r_2^2r_3^2
+\tilde{\gamma}_2^2r_1^2r_3^2.\eea The deformation parameters
$\tilde{\gamma}_i$ are related to $\gamma_i$ which appear in the
dual gauge theory as follows \bea\nn \tilde{\gamma}_i = 2\pi T
\gamma_i = \sqrt{\lambda} \gamma_i .\eea When
$\tilde{\gamma}_i=\tilde{\gamma}$ this becomes the supersymmetric
background of \cite{LM05}, and the deformation parameter $\gamma$
enters the $\mathcal{N}=1$ SYM superpotential in the following way
\bea\nn W\propto
tr\left(e^{i\pi\gamma}\Phi_1\Phi_2\Phi_3-e^{-i\pi\gamma}\Phi_1\Phi_3\Phi_2\right).\eea
This is the case we are going to consider here.

We restrict ourselves to the subspace $R_t\times S_{\gamma}^3$,
parameterize (see (\ref{roG})) \bea\nn r_1=\sin\theta , \h
r_2=\cos\theta ,\eea and use the ansatz \cite{KRT06} \bea\label{NRA}
&&t(\tau,\sigma)=\kappa\tau,\h \theta(\tau,\sigma)=\theta(\xi),\h
\phi_j(\tau,\sigma)=\omega_j\tau+f_j(\xi),\\ \nn
&&\xi=\alpha\sigma+\beta\tau,\h \kappa, \omega_j, \alpha,
\beta={\rm constants},\h j=1,2.\eea Then the string Lagrangian in
conformal gauge, on the $\gamma$-deformed three-sphere, can be
written as (prime is used for $d/d\xi$) \bea\nn
&&\mathcal{L}_\gamma=(\alpha^2-\beta^2)
\left[\theta'^2+G\sin^2\theta\left(f'_1-\frac{\beta\omega_1}{\alpha^2-\beta^2}\right)^2
+G\cos^2\theta\left(f'_2-\frac{\beta\omega_2}{\alpha^2-\beta^2}\right)^2
\right.
\\ \label{rl} &&-\left.\frac{\alpha^2}{(\alpha^2-\beta^2)^2}G
\left(\omega_1^2\sin^2\theta+\omega_2^2\cos^2\theta\right)
+2\alpha\tilde{\gamma}G\sin^2 \theta \cos^2 \theta \frac{\omega_2
f'_1-\omega_1 f'_2}{\alpha^2-\beta^2}\right], \eea where \bea\nn
G=\frac{1}{1+\tilde{\gamma}^2\sin^2 \theta \cos^2 \theta}.\eea

The equations of motion for $f_{1,2}$ following from (\ref{rl}) can
be integrated once to give \bea\label{fjs}  &&
f'_1=\frac{1}{\alpha^2-\beta^2} \left[\frac{C_1}{\sin^2 \theta}
+\beta\omega_1-\tilde{\gamma}\left(\alpha\omega_2-\tilde{\gamma}C_1\right)\cos^2\theta\right],
\\ \nn &&f'_2=\frac{1}{\alpha^2-\beta^2} \left[\frac{C_2}{\cos^2 \theta}
+\beta\omega_2+\tilde{\gamma}\left(\alpha\omega_1+\tilde{\gamma}C_2\right)\sin^2\theta\right]
,\eea where $C_{1,2}$ are integration constants.

Replacing (\ref{fjs}) into the Virasoro constraints one finds the
first integral $\theta'$ of the equation of motion for $\theta$ and
a relation among the parameters \bea\label{00r}
&&\theta'^2=\frac{1}{(\alpha^2-\beta^2)^2}
\Bigg[(\alpha^2+\beta^2)\kappa^2 -\frac{C_1^2}{\sin^2\theta}
-\frac{C_2^2}{\cos^2\theta}
\\ \nn &&-\left(\alpha\omega_1+\tilde{\gamma}C_2\right)^2\sin^2\theta
-\left(\alpha\omega_2-\tilde{\gamma}C_1\right)^2\cos^2\theta
\Bigg],
\\ \label{01r} && \omega_1C_1+\omega_2C_2+\beta\kappa^2=0.\eea

Now, we introduce the variable \bea\nn \chi=\cos^2\theta,\eea and
the parameters \bea\nn && v=-\frac{\beta}{\alpha},\h
u=\frac{\Omega_2}{\Omega_1},\h
W=\left(\frac{\kappa}{\Omega_1}\right)^2,\h
K=\frac{C_2}{\alpha\Omega_1},
\\ \nn &&\Omega_1=\omega_1\left(1+\tilde{\gamma}\frac{C_2}{\alpha\omega_1}\right), \h
\Omega_2=\omega_2\left(1-\tilde{\gamma}\frac{C_1}{\alpha\omega_2}\right)
.\eea By using them and (\ref{01r}), the three first integrals can
be rewritten as \bea\nn
&&f'_1=\frac{\Omega_1}{\alpha}\frac{1}{1-v^2} \left[\frac{v W-u
K}{1-\chi} -v(1-\tilde{\gamma}K)-\tilde{\gamma}u\chi\right],
\\ \label{tfi} &&f'_2=\frac{\Omega_1}{\alpha}\frac{1}{1-v^2} \left[\frac{K}{\chi}
-uv(1-\tilde{\gamma}K)-\tilde{\gamma}v^2W+\tilde{\gamma}(1-\chi)\right],
\\ \nn && \theta'
=\frac{\Omega_1}{\alpha}\frac{\sqrt{1-u^{2}}}{1-v^2}
\sqrt{\frac{(\chi_{p}-\chi)(\chi-\chi_{m})(\chi-\chi_{n})}{\chi(1-\chi)}},\eea
where \bea\nn &&\chi_p+\chi_m+\chi_n=\frac{2-(1+v^2)W-u^2}{1
-u^2},\\
\label{3eqs} &&\chi_p \chi_m+\chi_p \chi_n+\chi_m
\chi_n=\frac{1-(1+v^2)W+(v W-u K)^2-K^2}{1 -u^2},\\ \nn && \chi_p
\chi_m \chi_n=- \frac{K^2}{1 -u^2}.\eea We are interested in the
case of finite-size giant magnons, which corresponds to \bea\nn
0<\chi_{m}<\chi< \chi_{p}<1,\h \chi_{n}<0.\eea

Replacing (\ref{tfi}) and (\ref{3eqs}) in (\ref{rl}), we find the
final form of the Lagrangian to be (we set $\alpha=\Omega_1=1$ for
simplicity) \bea\nn &&\mathcal{L}_f=
-\frac{1}{1-v^2}\left[2-(1+v^2)W-2\tilde{\gamma}K
-2\left(1-\tilde{\gamma}K-u\left(u-\tilde{\gamma}uK+\tilde{\gamma}vW\right)\right)\chi\right].\eea

To obtain the finite-size effect on the three-point correlator, we
use (\ref{nsc}) and the explicit expression for the dilaton vertex
\bea \label{dv}
V^d=\left(Y_4+Y_5\right)^{-4}\left[z^{-2}\left(\p_+x_{m}\p_-x^{m}+\p_+z\p_-z\right)
+\p_+X_{k}\p_-X_{k}\right], \eea where \bea \nn
Y_4=\frac{1}{2z}\left(x^mx_m+z^2-1\right), \h
Y_5=\frac{1}{2z}\left(x^mx_m+z^2+1\right).\eea Here, $x_m$, $z$ are
coordinates on $AdS_5$, while $X_k$ are the coordinates on $S^5$.
This leads to \cite{Hernandez2,AB1105}
($i\tau=\tau_e$)\bea\label{c3d}
\mathcal{C}_3^{\tilde{\gamma}}=c_{\Delta}^{d}\int_{-\infty}^{\infty}\frac{d\tau_e}{\cosh^4(\kappa\tau_e)}
\int_{-L}^{L}d\sigma\left(\kappa^2 +\mathcal{L}_f\right).\eea
Performing the integrations in (\ref{c3d}), one finds
\bea\label{exact1}
&&\mathcal{C}_3^{\tilde{\gamma}}=\frac{16}{3}c_{\Delta}^{d}
\frac{1}{\sqrt{(1-u^2)W(\chi_p-\chi_n)}}\times \\ \nn
&&\left[\left((1-u^2)(1-\tilde{\gamma}K)-\tilde{\gamma}uvW\right)
\sqrt{\chi_p-\chi_n}\mathbf{E}(1-\epsilon)\right.
\\ \nn &&+\left.\left(\left(W(1-\tilde{\gamma}uv\chi_n)-(1-\tilde{\gamma}K)
\left(1-(1-u^2)\chi_n\right)\right) \mathbf{K}(1-\epsilon)\right)
\right],\eea where $\mathbf{K}(1-\epsilon)$ and
$\mathbf{E}(1-\epsilon)$ are the complete elliptic integrals of
first and second kind, and the following notation has been
introduced \bea\label{de}
\epsilon=\frac{\chi_{m}-\chi_{n}}{\chi_{p}-\chi_{n}}.\eea

This is our {\it exact} result for the normalized coefficient
$\mathcal{C}_3^{\tilde{\gamma}}$ in the three-point correlation
function, corresponding to the case when the heavy vertex operators
are {\it finite-size} dyonic giant magnons living on the
$\gamma$-deformed three-sphere.

For further purposes, let us also write down the exact expressions
for the conserved charges and the angular differences \bea\label{E}
&&\mathcal{E}\equiv \frac{2\pi E}{\sqrt{\lambda}}
=2\frac{(1-v^2)\sqrt{W}}{\sqrt{1-u^2}}\frac{
\mathbf{K}(1-\epsilon)}{\sqrt{\chi_{p}-\chi_{n}}},
\\ \label{J1} &&\mathcal{J}_1\equiv\frac{2\pi J_1}{\sqrt{\lambda}}=\frac{2}{\sqrt{1-u^2}}
\left[\frac{1-\chi_n-v\left(v W-u
K\right)}{\sqrt{\chi_{p}-\chi_{n}}}\mathbf{K}(1-\epsilon)-\sqrt{\chi_{p}-\chi_{n}}
\mathbf{E}(1-\epsilon)\right],\\ \label{J2}
&&\mathcal{J}_2\equiv\frac{2\pi
J_2}{\sqrt{\lambda}}=\frac{2}{\sqrt{1-u^2}} \left[\frac{u\chi_n-v
K}{\sqrt{\chi_{p}-\chi_{n}}} \mathbf{K}(1-\epsilon)
+u\sqrt{\chi_{p}-\chi_{n}} \mathbf{E}(1-\epsilon)\right],\eea  \bea
\label{p1}
&&p_1\equiv\Delta\phi_1=\phi_1(L)-\phi_1(-L)=\frac{2}{\sqrt{1-u^2}}
\\ \nn &&\times\Bigg\{\frac{v W-u
K}{(1-\chi_p)\sqrt{\chi_{p}-\chi_{n}}}
\Pi\left(-\frac{\chi_{p}-\chi_{m}}{1-\chi_{p}}\vert
1-\epsilon\right)-\left[v\left(1-\tilde{\gamma}K\right)+\tilde{\gamma}u\chi_{n}\right]
\frac{\mathbf{K}(1-\epsilon)}{\sqrt{\chi_{p}-\chi_{n}}}  \\
\nn &&-\tilde{\gamma}u\sqrt{\chi_{p}-\chi_{n}}
\mathbf{E}(1-\epsilon)\Bigg\},\eea \bea
\label{p2}&&p_2\equiv\Delta\phi_2=\phi_2(L)-\phi_2(-L)=\frac{2}{\sqrt{1-u^2}}
\\ \nn &&\times\Bigg\{\frac{K}{\chi_p\sqrt{\chi_{p}-\chi_{n}}}
\Pi\left(1-\frac{\chi_{m}}{\chi_{p}}\vert 1-\epsilon\right) -
\left[uv+\tilde{\gamma}v\left(v W-u
K\right)-\tilde{\gamma}\left(1-\chi_n\right)\right]
\frac{\mathbf{K}(1-\epsilon)}{\sqrt{\chi_{p}-\chi_{n}}}
\\ \nn &&-\tilde{\gamma}
\sqrt{\chi_{p}-\chi_{n}} \mathbf{E}(1-\epsilon)\Bigg\}.\eea Here,
$E$, $J_{1,2}$ are the string energy and angular momenta, while
$\phi_{1,2}$ are the isometry angles on the $\gamma$-deformed
three-sphere.

\setcounter{equation}{0}
\section{Small $\epsilon$ expansions}
For the case of the dilaton operator, the three-point function of
the SYM can be easily related to the conformal dimension of the
heavy operators. This corresponds to shift `t Hooft coupling
constant which is the overall coefficient of the Lagrangian
\cite{Costa}. This gives an important relation between the structure
constant and the conformal dimension as follows: \bea\label{rel1}
\mathcal{C}_3^{\tilde{\gamma}}=\frac{32\pi}{3}c_{\Delta}^{d}\sqrt{\lambda}\p_\lambda\Delta.
\eea We want to show here that this relation holds for the case of
finite-size giant magnons ($J_2=0$), assuming that $\Delta=E-J_1$,
and considering the limit $\epsilon\to 0$. To this end, we introduce
the expansions
\bea\nn
&&\chi_p=\chi_{p0}+\left(\chi_{p1}+\chi_{p2}\log(\epsilon)\right)\epsilon,
\\ \nn &&\chi_m=\chi_{m0}+\left(\chi_{m1}+\chi_{m2}\log(\epsilon)\right)\epsilon,
\\ \nn &&\chi_n=\chi_{n0}+\left(\chi_{n1}+\chi_{n2}\log(\epsilon)\right)\epsilon,
\\
\label{Dpars} &&v=v_0+\left(v_1+v_2\log(\epsilon)\right)\epsilon, \\
\nn &&u=u_0+\left(u_1+u_2\log(\epsilon)\right)\epsilon,
\\ \nn &&W=W_0+\left(W_1+W_2\log(\epsilon)\right)\epsilon,
\\ \nn &&K=K_0+\left(K_1+K_2\log(\epsilon)\right)\epsilon .\eea
A few comments are in order. To be able to reproduce the dispersion
relation for the infinite-size giant magnons, we set \bea\label{is}
\chi_{m0}=\chi_{n0}=K_0=0,\h W_0=1.\eea In addition, one can check
that if we keep the coefficients $\chi_{m2}$, $\chi_{n2}$, $W_2$ and
$K_2$ nonzero, the known leading correction to the giant magnon
energy-charge relation \cite{HS08} will be modified by a term
proportional to $\mathcal{J}_1^2$. That is why we choose
\bea\label{k2} \chi_{m2}= \chi_{n2}=W_2=K_2=0.\eea 
Finally, since we
are considering for simplicity giant magnons with one angular
momentum, we also set \bea\label{u0} u_0=0,\eea because the leading
term in the $\epsilon$-expansion of $\mathcal{J}_2$ is proportional
to $u_0$.

By replacing (\ref{Dpars}) in (\ref{3eqs}) and (\ref{de}), and
taking into account (\ref{is}), (\ref{k2}), (\ref{u0}), we obtain
\bea\label{chi} &&\chi_{p0}=1-v_0^2, \\
\nn &&\chi_{p1}= \frac{v_0}{1-v_0^2}
\Big[v_0\sqrt{(1-v_0^2)^4-4K_1^2(1-v_0^2)}-2(1-v_0^2)v_1 \Big ],\\
\nn &&\chi_{p2}=
-2v_0v_2 ,\\
\nn &&\chi_{m1}=
\frac{(1-v_0^2)^2+\sqrt{(1-v_0^2)^4-4K_1^2(1-v_0^2)}}
{2(1-v_0^2)}, \\
\nn &&\chi_{n1}=
-\frac{(1-v_0^2)^2-\sqrt{(1-v_0^2)^4-4K_1^2(1-v_0^2)}} {2(1-v_0^2)},
\\ \nn &&W_1=-\frac{\sqrt{(1-v_0^2)^4-4K_1^2(1-v_0^2)}}
{1-v_0^2}.\eea

The other parameters in (\ref{Dpars}) and (\ref{chi}) can be found
in the following way. First, we impose the conditions $J_2=0$ and
$p_1$ to be independent of $\epsilon$. This leads to four equations
with solution \bea\label{uv}
&&v_1=\frac{v_0\sqrt{(1-v_0^2)^4-4K_1^2(1-v_0^2)} \left(1-\log
16\right)}{4(1-v_0^2)}, \\ \nn &&v_2=
\frac{v_0\sqrt{(1-v_0^2)^4-4K_1^2(1-v_0^2)}}{4(1-v_0^2)}, \\ \nn
&&u_1=\frac{K_1v_0\log 4}{1-v_0^2}, \\ \nn &&u_2=
-\frac{K_1v_0}{2(1-v_0^2)},\eea where \bea\label{v0}
v_0=\cos\frac{p_1}{2}.\eea Next, to the leading order, the
expansions for $\mathcal{J}_1$ and $p_2=2\pi n_2$ $(n_2\in
\mathbb{Z})$ give \bea\label{ekf}
\epsilon=16\exp\left(-2-\frac{\mathcal{J}_1}{\sin\frac{p_1}{2}}\right),
\h K_1=\frac{1}{2}\sin^3\frac{p_1}{2}\sin\Phi, \h
\Phi=2\pi\left(n_2-\frac{\tilde{\gamma}}{\sqrt{\lambda}}J_1\right).\eea

Now, we consider the limit $\epsilon\to 0$ in the expression
(\ref{exact1}) for the structure constant in the 3-point correlation
function, by using (\ref{Dpars}), (\ref{is}), (\ref{k2}),
(\ref{u0}), (\ref{chi}), (\ref{uv}), and obtain \bea\label{C3se}
\mathcal{C}_3^{\tilde{\gamma}}&\approx &\frac{4}{3}c_{\Delta}^{d}
\frac{1}{(1-v_0^2)^{3/2}}\Bigg[4+4v_0^4\Big(1-\tilde{\gamma}K_1(1-\log
4)\ \epsilon\Big)\\ \nn
&&-v_0^2\left(8+\left(\sqrt{(1-v_0^2)^4-4K_1^2(1-v_0^2)}\ (1-\log
16)-8\tilde{\gamma}K_1(1-\log 4)\right)\epsilon\right)
\\ \nn
&&-\left(4\tilde{\gamma}K_1(1-\log
4)-\sqrt{(1-v_0^2)^4-4K_1^2(1-v_0^2)}\ (1-\log 256)\right)\epsilon
\\ \nn
&&-\left(v_0^2\sqrt{(1-v_0^2)^4-4K_1^2(1-v_0^2)}+2\tilde{\gamma}K_1(1-v_0^2)^2\right)
\epsilon\log\epsilon
\\ \nn
&&+\sqrt{(1-v_0^2)^4-4K_1^2(1-v_0^2)}\ \epsilon\log(16\
\epsilon)\Bigg] .\eea According to (\ref{v0}), (\ref{ekf}), the
above expression for $ \mathcal{C}_3^{\tilde{\gamma}}$ can be
rewritten in terms of $p_1$, $\mathcal{J}_1$, as \bea\label{pJ}
\mathcal{C}_3^{\tilde{\gamma}}&\approx & \frac{16}{3}c_{\Delta}^{d}
\sin\frac{p_1}{2}\left[1-4\sin^2\frac{p_1}{2}\left(\cos\Phi
+\mathcal{J}_1\csc\frac{p_1}{2}\cos\Phi
-\tilde{\gamma}\mathcal{J}_1\sin\Phi\right)
e^{-2-\frac{\mathcal{J}_1}{\sin\frac{p_1}{2}}}\right].\eea

In order to check if the equality (\ref{rel1}) holds for the present
case, let us now consider the dispersion relation of giant magnons
on $TsT$-transformed $AdS_5\times S^5$, including the leading
finite-size correction, which is known to be \cite{BF08,AB2010}
\bea\label{dr}
E-J_1=\frac{\sqrt{\lambda}}{\pi}\sin(p/2)\left[1-4\sin^2(p/2)\cos\Phi
\exp{\left(-2-\frac{2\pi
J_1}{\sqrt{\lambda}\sin(p/2)}\right)}\right].\eea Taking the
$\lambda$ derivative of (\ref{dr}), one finds \bea\label{der1}
\lambda\p_\lambda\Delta=\frac{\sqrt{\lambda}}{2\pi} \sin\frac{p}{2}
\left[1-4\sin^2\frac{p}{2}\left(\cos\Phi
+\mathcal{J}_1\csc\frac{p}{2}\cos\Phi
-\tilde{\gamma}\mathcal{J}_1\sin\Phi\right)
e^{-2-\frac{\mathcal{J}_1}{\sin\frac{p}{2}}}\right].\eea Identifying
$p\equiv p_1$, and comparing (\ref{pJ}) with (\ref{der1}), we see
that the equality (\ref{rel1}) is also valid for the
$\gamma$-deformed case.

\setcounter{equation}{0}
\section{Concluding Remarks}

In this note, we have derived the structure constant in the three-point
correlation function of two finite-size (dyonic) giant magnon string
states and a light dilaton state in the semiclassical approximation,
for the case of $\gamma$-deformed ($TsT$-transformed) $AdS_5\times
S^5$, dual to $\mathcal{N} = 1$ SYM, arising as
an exactly marginal deformation of $\mathcal{N} = 4$ SYM. 
We have confirmed our result by showing that the important relation between the
structure constant and the derivative of the conformal dimension
with respect to the t'Hooft coupling $\lambda$ still holds for the $\gamma$-deformed case. 
It will be interesting to consider correlation functions of other 
light operators or even all the heavy string states in the future.

\section*{Acknowledgements}
We would like to thank APCTP for hospitality where this work has been partially done.
This work was supported in part by WCU Grant No. R32-2008-000-101300
(C. A.) and DO 02-257 (P. B.).

\end{document}